\documentclass[12pt]{iopart}
\usepackage{amssymb}
\usepackage{epsfig}

\newcommand{\bbbone}{{\mathchoice {\rm 1\mskip -4mu l}{\rm 1\mskip
-4mu l}{\rm 1\mskip -4.5mu l}{\rm 1\mskip -5mu l}}}

\begin{document}

\title{Quantum coherent biomolecular energy transfer with spatially
correlated fluctuations}

\author{P Nalbach$^1$, J Eckel$^1$ and M Thorwart$^{1,2}$}
\address{$^1$School of Soft Matter Research, Freiburg Institute for
Advanced Studies (FRIAS), 
Albert-Ludwigs-Universit\"at Freiburg, Albertstra{\ss}e 19, 79104 Freiburg,
Germany }
\address{$^2$I.\ Institut f\"ur Theoretische Physik,  Universit\"at Hamburg,
Jungiusstra{\ss}e 9, 20355 Hamburg, Germany}

%\date{\today}

\begin{abstract}
We show that the quantum coherent transfer of excitations between
biomolecular chromophores is strongly influenced by spatial correlations
of the environmental fluctuations. The latter are due 
either to propagating environmental modes
or to local fluctuations with a finite localization length.
A simple toy model of a
single donor-acceptor pair with spatially separated chromophore
sites allows to investigate the
influence of these spatial correlations on the quantum coherent
excitation transfer.
The sound velocity of the solvent determines the wave lengths of the
environmental
modes, which, in turn, has to be compared to the spatial distance of the
chromophore sites. When the wave length exceeds the distance 
between donor and acceptor site, 
we find strong suppression of decoherence.
In addition, we consider two spatially separated
 donor-acceptor pairs under the influence of propagating environmental
modes.
Depending on their wave lengths fixed by the sound velocity of the
solvent material,
the spatial range of correlations may extend over typical interpair
distances, which
can lead to an increase of the decohering influence of the solvent.
Surprisingly, this effect is counteracted by increasing temperature.
\end{abstract}

%03.65.Yz 	Decoherence; open systems; quantum statistical methods 
%05.60.Gg 	Quantum transport 
%71.35.-y 	Excitons and related phenomena
%87.15.ht 	Ultrafast dynamics; charge transfer
\submitto{\NJP}

\section{Introduction}

The photosynthetic conversion of physical energy of sunlight into its
chemical form suitable for cellular processes involves many physical
and chemical mechanisms~\cite{Bio1,Bio2}. Photosynthesis starts with
the absorption of a photon by a light-harvesting pigment forming an
exciton, followed by the transfer of the exciton to the reaction
center, where charge separation is initiated. It is the nature of this
transfer process which is presently in the focus of intense research.
Recent experiments \cite{BioEn2007,Bio5} provided evidence that an
incoherent hopping model seems not to be sufficient to
describe long-lasting beating signals in a two-dimensional Fourier
transform electronic spectrum \cite{Brixner} 
recorded from green sulfur bacteria such as {\it Chlorobium tepidum}.
Here, the energy transfer between the main chlorosome antenna and the
reaction centers is mediated by the Fenna-Matthews-Olson (FMO)
protein~\cite{Bio3,Bio4,Rienk} which 
contains bacteriochlorophyll (BChl) molecules. The FMO protein is a
trimer made of identical subunits, each of which contains seven BChl
molecules and no carotenoids. Due to its small size, it represents an
important model system for photosynthetic energy transfer and has been
extensively studied experimentally and theoretically. The observed
\cite{BioEn2007} long lived electronic coherence lasted up to time
scales comparable to the time scale of the energy transport. The
experiments were performed at low temperature $T=77$~K and clearly
suggest that the exciton moves coherently through the FMO complex
rather than by incoherent hopping. Similarly, Lee et al.~\cite{Bio5}
found coherent beating signals at low temperatures in a two-color
electronic coherence photon echo experiment. It allows to directly
probe electronic coherences by mixing of the bacteriopheophytin and
accessory bacteriochlorophyll excited states in the reaction center of
the purple bacterium Rhodobacter sphaeroides. These measurements were
performed at 77 K and at 180 K. The coherence beatings in these
measurements can only be explained \cite{Fleming09a} by a strong
correlation between protein-induced fluctuations in the transition
energies of neighboring chromophores, leading to the conclusion that
protein-correlated environments in fact preserve and support
electronic coherence in photosynthetic complexes. 

Recently, an ultrafast polarization experiment \cite{Scholes} has
revealed quantum coherent intrachain (but not interchain) electronic
energy transfer in conjugated polymers with different chain
conformations as model multichromophoric systems at room temperature.
The data suggest that chemical donor-acceptor bonds help to correlate
dephasing perturbations. By introducing an angle-resolved  coherent
optical wave mixing technique, the quantum beating signals between
coherently coupled electronic transitions in the light-harvesting
complex of purple bacteria are directly observed \cite{Cogdell09}.
Also in the light-harvesting proteins of cryptophyte marine algae,
quantum coherent couplings have been identified by exceptionally
long-lasting excitation oscillations even at ambient temperature
\cite{Scholes10}. 

The consequences of these seminal experiments on long-lived
electronic coherences are immediate: the excitation can move rapidly
and reversibly in space, allowing for a very efficient search for an
energetic minimum in photosynthesis. The subsequent trapping of the
excitation, however, must be optimized by properly adjusted
environmental fluctuations. In view of potential applications, the
recent progress helps to understand the design principle of
photosynthetic complexes \cite{Fleming09c} and to exploit the
near-unity efficiency of energy transfer which is believed to result
from the constructive interplay of quantum coherence and slow, 
spatially correlated environmental fluctuations. This could open the
door to efficient future artificial light-harvesting complexes finding
applications in optimized organic solar cells. The combination of
optimized exciton trapping \cite{Silbey} with powerful quantum
coherent adaptive control schemes \cite{Motzkus02,Motzkus05,Motzkus08}
could, in addition, allow to exploit quantum effects to direct the
outcome of photochemical processes. Effects beyond the lowest-order
F\"orster treatment, such as the failure of the point-dipole
approximation  and the ensuing solvent screening and the 
sharing of common bath modes have been considered in a generalized
F\"orster theory \cite{Silbey09}.  

%% Theory:

Theoretically, photosynthetic energy transfer processes in
light-harvesting complexes are often discussed in simplified
low-dimensional models describing a few individual chromophores
which mutually interact by dipolar couplings and which are 
exposed to the fluctuations of the polar solvent molecules and the
protein host
\cite{Bio1,Bio2,SpiBoLe1987,SpiBoWeiss,BioGi2005,BioGi2006,
BioGi2007}. 
Two limiting cases are commonly considered: (i) When the dipole
coupling between the chromophores is weak in comparison to the
coupling of the chromophore to environmental fluctuations, the
excitons are considered to be localized at the chromophore sites. The
weak
electronic coupling can then be treated perturbatively, resulting in
an incoherent hopping dynamics described by the standard F\"orster
theory~\cite{Bio6}. (ii) In the opposite limit of weak coupling to the
environmental fluctuations, standard perturbative quantum master
equations are used 
\cite{Aspuru08,Aspuru09a,Aspuru09b,Alexandra08,Alexandra09a,
Alexandra09b,Plenio08,Plenio09,Plenio10a,Plenio10b,Plenio10c},
resulting 
in a damped coherent dynamics for the exciton transfer. They naturally
are based on the Markovian approximation, which renders the time
evolution 
memoryless and allows for a straightforward numerical solution. 
These types of master equations are appropriate in the case of a clear
separation of time scales, i.e., 
when the time scale on which the environmental fluctuations occur is
much smaller than  that on which the system dynamics evolves.
Formally, this is captured by the requirement that the bath
reorganization energy \cite{SpiBoWeiss} is much larger than the typical
system energy. This condition is typically not fulfilled for the
energy transfer dynamics in biomolecular light-harvesting complexes in
a protein-solvent environment
\cite{BioGi2005,BioGi2006,BioGi2007,Valkunas}. Here, in addition to 
the rather slow polarization fluctuations of the polar solvent
molecules, the protein cage acts also as a frequency filter which 
particularly shapes the frequency distribution of the environmental
modes. A
similar effect arises for electronic spin qubits in semiconductor
quantum dots \cite{Bio10a} and in donor-based charge qubit crystal
systems
\cite{Bio10b}. Here, the designed geometrical shape of the system
structures the acoustic phonon spectrum and gives rise to strong
non-Markovian effects. In the context of excitonic energy transfer, it
has recently been shown~\cite{Bio7,BioTh2008,Bio8} 
that under realistic biomolecular circumstances, time-local master
equations become increasingly unreliable when both time scales become
comparable. In addition to the violation of the Markovian assumption,
the coupling between chromophores and environment cannot be
considered as weak enough to allow for a lowest order perturbative
treatment \cite{BioGi2005,BioGi2006,BioGi2007}. It was recently shown
that the latter condition renders any weak coupling approach
questionable~\cite{Bio8}. Non-Markovian approaches have been employed
beyond a lowest-order treatment 
\cite{Bio7} and by coupling each chromophore to a single damped
harmonic mode \cite{Plenio10c}, whose existence, however,  was not
further motivated. Alternatively, numerically exact simulations of the
real-time dynamics of quantum coherent energy transfer 
under realistic conditions have been carried out
\cite{BioTh2008,Bio8,BioTh2009}
 by employing the quasiadiabatic propagator path-integral (QUAPI)
\cite{QUAPI1,QUAPI2,Bio11}. 
By this, it has been shown that the rather slow polarization
fluctuations are one possibility to enhance quantum coherence 
in the transfer processes. The coupling of two chromophore pairs to
the common slowly fluctuating modes even allows to entangle two 
excitonic pairs over surprisingly long times even at room temperature
\cite{BioTh2008}. However, when the fluctuations are fast, no
entanglement is created even when the two pairs couple to the same
modes. The possibility of entanglement and the role of non-Markovian
contributions in biomolecular complexes have also been re-addressed in
 recent works \cite{Plenio10a,Plenio10b,Buch10}.  
Slow fluctuations have also been treated \cite{Mukamel10} by 
different variants of cumulant expansion techniques and by statistical
averaging approaches over static disorder in sum-over-eigenstates
approaches. Recently, the standard Redfield equations, which are valid
in the weak-coupling regime, have been extended by generalizing the
Redfield relaxation tensor on the basis of the Lindblad quantum master
equation \cite{Mukamel09}. This technique goes beyond the secular
approximation and thus can include effects of stronger coupling.
However, the approach is still memoryless and leads to time-local
evolution equations.

%% Our work

Despite the fact that the experimental coherence
beatings~\cite{Fleming09a} could only be explained including strong
correlations between protein-induced fluctuations in the transition
energies of neighboring chromophores, the influence of these
correlations received little attention in the theoretical
investigations. Nazir~\cite{Nazir09} investigated the influence of
correlated fluctuations on a donor-acceptor system for strong system
bath coupling. Correlations in a super-Ohmic bath are found to 
suppress the crossover to incoherent dynamics at high temperatures, 
which is in line with the common expectation \cite{SpiBoWeiss} that a
super-Ohmic bath naturally provides only weakened influence of the
fluctuations on the system.  Similar effects would be expected for
correlations in the environment of the chromophores whose spectrum is
typically assumed to be Ohmic. Fassiolo et al.~\cite{Alexandra09a}
discussed the influence of correlations on the trapping probability in
a ring of chromophores within a Lindblad master equation approach.
They find that the correlations between environmental fluctuations
allows to tune the trapping probability.

From a condensed matter point of view, environmental fluctuations are
normal modes of the bulk material and thus are typically phonons which
propagate through the material. Accordingly they couple to all
chromophores with amplitude differences determined by the phase
differences due to finite times the modes need to propagate from one
chromophore site to the next. However, rattling of side chains of
macromolecules in these highly disordered protein environments might
well be viewed as a localized excitation. In most existing studies it
is assumed that each site of a multichromophoric array is coupled to
its local environment. Including in these approaches spatial
correlations could be achieved by assuming a finite localization
length of the excitations.

In the next section we discuss the influence of spatially correlated
environmental fluctuations on a donor-acceptor model using the
numerical exact quasi-adiabatic path integral propagator approach
which allows us to treat realistic strong couplings and slow
environments. Although being a clear oversimplification to realistic
exciton
transport the donor-acceptor model serves as a toy model to study the
influence of spatial correlations and the difference between
propagating and localized modes on quantum coherence in detail. In the
third section we will discuss how spatial correlations influence two
donor-acceptor pairs which are initially uncoupled. Depending on the
distance $r_{da}$ between donor and acceptor and the distance $r$
between the two donor-acceptor pairs spatially correlated fluctuations
increase or decrease the decay rates of the coherent dynamics. Finally
we discuss and summarize our results.

\section{Spatial environmental correlations in a single chromophore 
pair}
\label{single}
\subsection{Model}

The simplest way to model a single chromophore (or pigment) is by
describing it as a quantum two-level system consisting of a ground and
an excited state, which are separated by the energy gap $\epsilon$.
When the electron is in the excited state, it is 
localized by its attractive interaction with the hole it left. This
dipole electron-hole configuration forms an exciton. A formal
description can be given 
in terms of the Pauli matrix $\tau_z$. Environmental
fluctuations will cause transitions between the ground and the excited
state and will add a fluctuating energy. Experimentally it is known
that the recombination time is of the order of nanoseconds, whereas the
complete energy transfer through the complex is of the order of
picoseconds. Thus the environmental fluctuations causing recombination
are negligible. Describing the fluctuations by harmonic oscillators,
which couple linearly to the chromophore, results in the independent
boson model for a single
chromophore
\begin{equation}
 H \,=\, \epsilon\frac{\tau_z}{2} \,- |e\rangle\langle
e|\sum_{\bf k}\lambda_{\bf k}({\bf r}) q_{\bf k} \,+\frac{1}{2}\sum_{\bf k}
\left(p_{\bf k}^2+\omega_{\bf k}^2q_{\bf k}^2 \right) 
\end{equation}
where we introduced the position and momentum operators, $q_{\bf k}$ and
$p_{\bf k}$, of the mode with wave vector ${\bf k}$ and its coupling
$\lambda_{\bf k}({\bf r})$ to the chromophore which depends on the amplitude
of the fluctuation at the position ${\bf r}$ of the chromophore. We
explicitely coupled the environmental fluctuations only to the excited
state, $|e\rangle$, which expresses the fact that the electronic
ground state energy is defined by including all vibrational
equilibrium energies. We fixed $\hbar=k_B=1$ which we keep below.

We are not aiming at a complete description of exciton transfer
dynamics in complexes like FMO but are merely interested in the
question how spatially correlated environmental fluctuations influence
the transfer process between two excitonic sites. 
For the sake of simplicity, we restrict the model under consideration
to two chromophore sites (acceptor and donor at ${\bf r}_{a/d}$) which
contain a single exciton. We do not consider different site energies
(i.e., $\epsilon_a=\epsilon_d$) 
and are thus lead to the donor-acceptor Hamiltonian
\begin{equation}
\label{Hda} H_{\rm da} = \frac{1}{2}\Delta \left\{ |d\rangle\langle
a| \,+|a\rangle\langle d| \right\} \,+ \sum_{i=a/d}|i\rangle\langle i|
\sum_{\bf k}\lambda_{\bf k}({\bf r}_i) q_{\bf k} \,+\frac{1}{2}\sum_{\bf k}
\left(p_{\bf k}^2+\omega_{\bf k}^2q_{\bf k}^2 \right) \, 
\end{equation}
of a single chromophore pair. 
The state $|d\rangle$ ($|a\rangle$) denotes the exciton to be at the
donor (acceptor) and $\Delta$ is the respective dipole coupling 
matrix element \cite{BioGi2005}.\\[3mm]
{\it Comparison of the donor-acceptor model with the spin-boson
model}\\[3mm]
The donor-acceptor Hamiltonian, Eq.~(\ref{Hda}), can easily be
transformed into a Hamiltonian which is closer to the widely
studied spin-boson model~\cite{SpiBoLe1987,SpiBoWeiss}
\begin{eqnarray}
\label{spibo} 
H_{\rm da} &=& \Delta\, \frac{\sigma_x}{2} \,+
\frac{\sigma_z}{2} \sum_{\bf k} \left\{\lambda_{\bf k}({\bf r}_2) - \lambda_{\bf k}({\bf
r}_1) \right\} q_{\bf k}  \,+ \frac{\bbbone}{2} \sum_{\bf k}
\left\{\lambda_{\bf k}({\bf r}_2) + \lambda_{\bf k}({\bf r}_1) \right\} q_{\bf k}  \nonumber \\ 
& &  +\frac{1}{2}\sum_{\bf k} \left(p_{\bf k}^2+\omega_{\bf k}^2q_{\bf k}^2 \right) 
\end{eqnarray}
by introducing the Pauli matrices
$\{\bbbone,\sigma_x,\sigma_y,\sigma_z\}$ with
$\sigma_x=|d\rangle\langle a| +|a\rangle\langle d|$ and
$\sigma_z=|d\rangle\langle d| -|a\rangle\langle a|$.

One difference is given by the term proportional to $\lambda_{\bf k}({\bf
r}_1)+\lambda_{\bf k}({\bf r}_2)$ which couples to the identity operator
$\bbbone$ of the donor-acceptor system thus causing fluctuations of
the reference energy of the donor-acceptor system. Accordingly
it is irrelevant for its dynamics. 
However, this term modifies the bath modes by shifting their 
zero-point energies and thus changes their thermal equilibrium state.
In the spirit of dissipative quantum dynamics, the treatment of
system-bath problems typically rely on the assumption that the bath is
only weakly influenced by the coupling to the system itself and thus
the mentioned effects should not affect the dissipative dynamics
of the donor-acceptor system. However, at strong coupling or for a
slow bath with cut-off frequency $\omega_c\lesssim\Delta$ the validity
of these assumptions is questionable~\cite{Reinhold}. Thus the
dynamics generated by our donor-acceptor Hamiltonian in Eq.\
(\ref{Hda}) differs from the
standard spin-boson problem in three aspects. First, the thermal
equilibrium state to which the total system is driven is
different and, second, the factorized initial conditions for both
cases reflect two  different initial conditions. Third,  in the 
spin-boson model, a single bath is coupled to the system whereas in
the donor-acceptor Hamiltonian, two baths are
coupled to the system states. Thus, even when both baths are
mutually uncorrelated, the resulting rates for the donor-acceptor
system are twice as large compared to the spin-boson model (assuming
that all baths are coupled with equal strength). This has to be taken
into account when comparing results from both approaches.

\subsection{The quasiadiabatic propagator
path integral for the multi-bath case}

The dynamics of the donor-acceptor is characterized by the time
evolution of the reduced density matrix $\rho(t)$, which is obtained
after tracing out the environmental (or bath) degrees of freedom,
i.e.,
\begin{equation}  
\rho(t) \,=\, {\rm Tr} \left\{{U(t,0)W(0)U^{-1}(t,0)} \right\}_B  \\
\end{equation}
and
\begin{equation}   
U(t,0) \,=\, {\cal T} \exp\left\{ -\frac{i}{\hbar} \int_0^t ds H_{da}
\right\} \, . 
\end{equation}
Here, $U(t,0)$ denotes the propagator of the full system plus
bath and ${\cal T}$ denotes the time-ordering operator. $W(0)$ is the
total density operator at initial time set at $t=0$. We assume
standard factorizing initial conditions~\cite{SpiBoWeiss}, i.e. 
$W(0)\propto\rho(0)\exp(-H_B/T)$, where the bath with the Hamiltonian 
$H_B=\frac{1}{2}\sum_k \left(p_{k}^2+\omega_k^2q_{k}^2 \right)$ is at
thermal equilibrium
at temperature $T$ and the system is prepared according
to $\rho(0)$. Throughout this work, we always start with the exciton 
at the donor site, i.e.,  $\rho(0)=|d\rangle\langle d|$.

We calculate $\rho (t)$ using the numerically exact quasiadiabatic
propagator path-integral (QUAPI)~\cite{QUAPI1,QUAPI2,Bio11} scheme.
For details of the iterative technique, we refer to previous works
\cite{QUAPI1,QUAPI2,Bio11}. In brief, the algorithm is based on a
symmetric Trotter splitting 
of the short-time propagator ${\cal K}(t_{k + 1}, t_k)$ for the full
Hamiltonian into a part depending on the system Hamiltonian and a part
involving the bath and the coupling term. The short-time propagator
describes time evolution over a Trotter time slice $\delta t$. This
splitting is by construction exact in the limit $\delta t \to 0$ but
introduces a finite Trotter error for a finite time increment, 
which has to be eliminated by choosing $\delta t$ small enough such
that convergence is achieved. On the other side, the bath degrees of
freedom generate correlations being non-local in time. For any finite
temperature, these correlations decay exponentially fast at asymptotic
times, thereby setting the associated memory time scale. QUAPI now
defines an object called the reduced 
density tensor, which lives on this memory time window and establishes
an iteration scheme in order to extract the time evolution of this
object. Within the memory time window, all correlations are included
exactly over the finite memory time  $\tau_{\rm mem} = K \delta t$ and
can safely be neglected for times beyond $\tau_{\rm mem}$. 
Then, the memory parameter $K$ has to be increased, until convergence
is found. The two strategies to achieve convergence are naturally
countercurrent, but nevertheless 
convergent results can be obtained in a wide range of parameters.

For the purpose of this work, we have to extend the standard
formulation of QUAPI which only includes the coupling to 
one bath. The entire influence of a single bath coupled via the
operator $\hat{s}$ to the donor-acceptor system is described in terms of the real-time 
path-integral formulation by the influence functional
\begin{equation}\label{InfluenceFunc}
 I(\{s_i^+,s_i^-\};\delta t) \,=\, 
\exp\left\{ -\frac{1}{\hbar}\sum_{i=0}^N\sum_{i'=0}^i \,
[s_i^+-s^-_i] 
\left[\eta_{ii'}s_{i'}^+-\eta_{ii'}^* s^-_{i'}\right] \right\} 
\end{equation}
where the path segments $s_i^\pm$ associated to a Trotter time slice 
$i$ given as interval 
$[(i-\frac{1}{2})\delta t,(i+\frac{1}{2})\delta t]$ (with total time $t=N\delta t$) 
are assumed to have constant values over a single time slice. The number 
of path segments within a Trotter time slice is given by the dimension 
of the Hilbert space in which the system-bath coupling operator lives. 
The superscript $\pm$ denotes the propagation direction forward or backward in 
time since we work with density operators. The total path integration over 
all paths $s^\pm (t')$ has to be performed as the discrete sum over 
all configurations $\{s_i^+,s_i^-\}$ of paths segments between initial and final 
time. The time-discrete 
bath correlators $\eta_{ii'}$ are defined in Ref.~\cite{QUAPI1} and the 
superscript $^*$ denotes the complex conjugate.

Multiple independent baths, $H_{B\alpha}$, which couple to system operators
$\hat{s}_\alpha$ will simply cause a product of influence functionals
since each bath acts separately as described above. Thus,  the
total influence functional assumes the form 
\begin{equation}\label{InfluenceFunc2}
 I(\{s_{i,\alpha}^+,s_{i,\alpha}^-\};\delta t) \,=\, 
\exp\left\{ -\frac{1}{\hbar}\sum_\alpha\sum_{i=0}^N\sum_{i'=0}^i \,
[s_{i,\alpha}^+-s_{i,\alpha}^-] 
\left[\eta_{ii'}^{(\alpha\alpha)}s_{i',\alpha}^+-\eta_{ii'}^{*
(\alpha\alpha)} s_{i',\alpha}^-\right] \right\} 
\end{equation}
Here we denoted the bath correlators $\eta_{ii'}^{(\alpha\alpha)}$
with the additional superscripts since for differing baths the
correlators will differ.

The question whether the environmental fluctuations act {\it locally}
or in a correlated manner can be tackled by the following extension. 
Local fluctuations couple to the donor and to the acceptor separately and 
independently. This implies that 
Eq.\ (\ref{InfluenceFunc2}) describes {\em all\/} effects due to
environmental fluctuations. If, however, the fluctuations are caused
by extended waves, like phonon modes or if the fluctuations rattling the
donor can at least partially still be felt at the acceptor site, then the
fluctuations at the various sites are no longer independent and spatial 
correlations have to be taken into account. 
Hence, Eq.\ (\ref{InfluenceFunc2}) has to be generalized to
\begin{equation}\label{InfluenceFunc3}
 I(\{s_{i,\alpha}^+,s_{i,\beta}^-\};\delta t) \,=\, 
\exp\left\{ -\frac{1}{\hbar}\sum_{\alpha,\beta}\sum_{i=0}^N\sum_{i'=0}^i
\, [s_{i,\alpha}^+-s_{i,\alpha}^-] 
\left[\eta_{ii'}^{(\alpha\beta)}s_{i',\beta}^+-\eta_{ii'}^{*
(\alpha\beta)} s_{i',\beta}^-\right] \right\} 
\end{equation}
where $\eta_{ii'}^{(\alpha\beta)}$ are the mixed bath correlators
expressing the correlations of the fluctuations acting at operator
$\hat{s}_\alpha$ and $\hat{s}_\beta$.
The detailed numerical evaluation of the influence functional in the 
extended QUAPI scheme becomes more involved,
but the general procedure is not affected by this extension.

\subsection{Correlated environmental fluctuations at different sites}

In a crystal, environmental fluctuations acting, e.g., on electrons, 
are generated by vibrations of the lattice atoms and are the well-known 
phonons. Phonons are also present in disordered
media (condensed, soft or fluid). In the sense of propagating modes 
of the host material which evolve with time through the medium, 
they are commonly limited to the low energy sector or, more specifically, to
energies associated to wave lengths on which the material appears
homogeneous. Once the wave length becomes smaller than the disorder
length 
scale, 
the modes can generally be thought of as localized fluctuations. However,
their localization length (or radius) is still connected to the wave
length of the mode and accordingly even a {\it localized} mode extends
over some finite spatial range. The same picture holds for dipolar 
fluctuations in solvents and vibrations of charged 
macromolecular side chains forming
the bio-environment of light harvesting complexes.
\begin{figure}[t]
\begin{center}
\epsfig{file=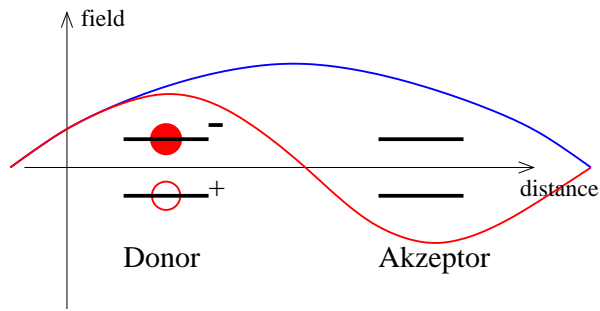,width=8cm}
\end{center}
\caption{\label{fig1} Donor and acceptor with two environmental field
with different wave length are illustrated. The donor holds an
electron in the excited state which forms with the hole left in the
ground state an exciton having a finite
electric dipole moment.}
\end{figure}
Excitons, as coupled electron hole pairs, have an electric dipole moment
$\vec\mu$ which couples to the electric field $\vec{E}({\bf r})$ at the
exciton position ${\bf r}$ generated by the environmental dipolar
fluctuations (as illustrated in Fig.~\ref{fig1}). This results in the
interaction energy $H_{SB}=\vec\mu\cdot \vec{E}({\bf r})$.
For simplicity, we ignore the angular dependence in the following which
only cause correction factors of the order of one~\cite{Peter}. Within
the donor-acceptor model the exciton dipole moment is described
by $\mu=|\vec\mu|=\mu_0|i\rangle\langle i|$ where we furthermore have assumed
that the dipole moments at each chromophore site are the same (again
neglecting factors of the order one due to angular dependencies).
The electric field is proportional to the amplitude of the propagating
normal modes of the medium, $E({\bf r})={\rm sgn}\{\vec{E}({\bf
r})\}\cdot |\vec{E}({\bf r})|=E_0(1/\sqrt{N})\sum_{\bf k} q_{\bf k} e^{i
{\bf kr}}$ finally leading to the interaction Hamiltonian in
Eq.~(\ref{Hda}) with $\lambda_{\bf k}({\bf r})=(\mu_0E_0/\sqrt{N})e^{i
{\bf kr}}$.
Similar ideas have recently been used to understand the phonon influence
on double quantum dot charge qubits~\cite{Bio10a,Bio10b} or tunneling
defects~\cite{Peter}.

For propagating modes in three spatial dimensions the spectral function
of intersite fluctuations between chromophores $i$ and $j$ becomes
\begin{eqnarray} J_{ij}(\omega) &=& \sum_{\bf k} \frac{\lambda_{\bf k} ({\bf r}_i)
\lambda_{-{{\bf k}}}({\bf r}_j) }{2\omega_{\bf
k}}\delta(\omega-\omega_{\bf k})
\,=\, 2\alpha\omega e^{-\omega / \omega_c} \frac{\sin(\omega
t_0)}{\omega t_0} \nonumber \\ 
& &  \quad\mbox{with}\quad t_0 \,=\, \frac{r_{ij}}{v}
\end{eqnarray}
with the sound velocity $v$ (assuming linear dispersion $\omega_{\bf
k}=vk$ and $k=|{\bf k}|$), the distance $r_{ij}=|{\bf r}_i-{\bf r}_j|$
between site $i$ and $j$, coupling strength $\alpha$ and upper cut-off
$\omega_c$ using an exponential form for the cut-off function. For the
small cut-off frequencies typical for biomolecular environments
($\omega_c\simeq\Delta$) the cut-off function will modify quantitatively
but not qualitatively the results. However, no detailed information
about
the specific cut-off functions for biomolecular environments is
available in the literature. The (on-site) spectrum is Ohmic (linear in
$\omega$) for the FMO complex~\cite{BioCh2005,BioGi2007}.
Linear dispersion for the normal modes is a strong assumption and it is
not clear if the simple Debye picture holds in biological soft matter up
to energies $\Delta$.
For on-site fluctuations at site $i$ the spectral function simplifies
\begin{equation} J_{ii}(\omega) \,=\, \sum_{\bf k} \frac{|\lambda_{\bf k}({\bf
r}_i)|^2}{2\omega_{\bf k}}\delta(\omega-\omega_{\bf k})
\,=\, 2\alpha\omega e^{-\omega / \omega_c} \, . \label{onsite}
\end{equation}

Alternatively one might consider localized environmental fluctuations
with localization length $\xi$, which is taken to be 
 independent of the mode energy $\omega$, resulting in the spectral 
density 
\begin{equation} J_{ij}^{\rm loc}(\omega) 
\,=\, 2\alpha\omega e^{-\omega / \omega_c} e^{-r_{ij}/\xi}
\end{equation}
for the intersite spectrum  whereas the on-site spectrum, 
Eq.\ (\ref{onsite}), is unaltered.

The intersite spectrum $J_{ij}^{\rm loc}(\omega)$ vanishes for sites far
apart $r_{ij}\gg\xi$ and the fluctuations at the donor and the acceptor 
sites  are
uncorrelated. The intersite spectrum becomes identical to the on-site
spectrum for close sites $r_{ij}\ll\xi$. In the later case, the
environmental fluctuations of both sites are fully correlated and thus
actually identical. As illustrated in Fig.~\ref{fig1} by the blue line,
both excitons then ``see'' the same electric field, which, in turn, 
only modifies the total
energy but not the energy difference between donor and acceptor.
Accordingly, these fully correlated fluctuations cannot influence the
dynamics of the donor-acceptor system.
\begin{figure}[t]
\begin{center}
\epsfig{file= 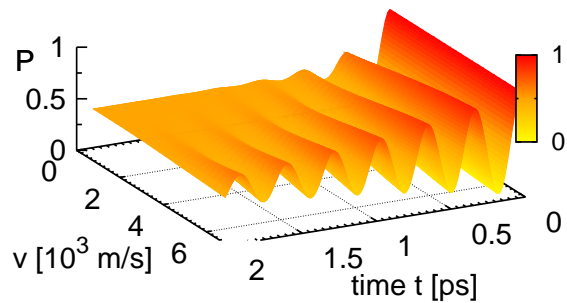,width=8cm}
\end{center}
\caption{\label{fig2} Occupation difference of donor and acceptor versus
time and speed of sound for $\omega_c=\Delta=106$~cm$^{-1}$, $T=152$~K,
$\alpha=0.08$ and $r_{da}=3.8$~\AA~assuming propagating environmental
modes.}
\end{figure}

For the case of propagating modes, qualitatively the same holds.
When $\Delta t_0\gg 1$ all modes with $\omega\ge\Delta$ will not
contribute to the intersite spectral function. At least at weak coupling 
($\alpha \ll 1$), 
mainly the modes resonant with the tunneling splitting are relevant and
thus we expect the intersite spectrum to be irrelevant. The environment
acts as two independent baths at each chromophore site, 
as usually assumed in the literature.
When the shortest wave length $\lambda_c$ in the spectrum 
is larger than the distance between donor and acceptor, 
we have that $\omega_ct_0=(r_{ij}/\lambda_c)\ll 1$. 
Then, the environmental fluctuations are fully
correlated between the sites  and thus 
do not affect the donor-acceptor system.

For the donor-acceptor system we can transform the Hamiltonian as
discussed in Eq.~(\ref{spibo}) and then define a single effective
spectral function (as for a spin-boson
problem~\cite{SpiBoWeiss,SpiBoLe1987}) resulting in
\begin{equation} J_{\rm eff}(\omega) \,=\, 4\alpha\omega e^{-\omega / \omega_c}
\left( 1- \frac{\sin(\omega t_0)}{\omega t_0} \right) \, .
\end{equation}
In the limit $\omega_c t_0\ll 1$, we get
\begin{equation} J_{\rm eff}(\omega) \,\simeq\, 4\alpha t_0^2\omega^3 e^{-\omega /
\omega_c}
\end{equation}
which is of super-Ohmic form. Super-Ohmic environmental fluctuations can
neither cause overdamping (except at large temperatures) nor localization, 
in clear qualitative contrast to pure Ohmic fluctuations. 
This is a drastic qualitative effect which spatially 
correlated environmental fluctuations cause 
on coherent exciton transfer.

\subsection{Dynamics of a single transfer step}

Having determined the reduced density matrix by QUAPI, we can evaluate the 
occupation difference $P(t)=\langle \sigma_z \rangle$ between donor and
acceptor. 
In Fig.~\ref{fig2}, $P(t)$ is plotted over time versus sound velocity 
$v$ with which the modes are assumed to propagate. 
We have chosen all parameter to match rather closely the properties of 
 chromophores in the FMO complex~\cite{BioCh2005}. We have used 
$\Delta=106$~cm$^{-1}$ as tunneling element
 which corresponds to the largest coupling in
the FMO complex~\cite{BioCh2005} between chromophore $1$ and $2$ but
have neglected the energy difference between the two sites. The distance
between site $1$ and $2$ in the FMO complex of {\it Chlorobium
tepidum}~\cite{Bio4} is $r_{12}=3.8$~\AA~which are the closest two
chromophores. Site $2$ and $7$ are maximally apart, $r_{27}=11.3$~\AA.
The bath cut-off frequency varies in the
literature~\cite{BioCh2005,BioAd2006} between $\omega_c=32$~cm$^{-1}$
and $150$~cm$^{-1}$. To be specific, we choose $\omega_c=106$~cm$^{-1}$ and
temperature $T=152$~K$=\Delta/k_B$.

We find that quantum coherent oscillations occur which decay within about
1~ps for the smallest value of the sound velocity of several hundred m/s. The
environmental fluctuations are uncorrelated in this case. 
For larger sound velocities,
meaning increasing correlations of the fluctuations, the decay slows down
considerably as expected since the wave length of the modes causing
decoherence becomes larger than the distance between the chromophores
and thus cannot harm coherence any longer. We are not aware of
experimental data regarding the precise values of sound velocities for the biological
embedding materials of the FMO complexes. As a guide we might use
the sound velocity of water, $v\simeq 1500$~m/s, which falls into the
range of our plot. When frozen to ice, as in the low temperature
experiments at $77$~K or $180$~K, one finds that $v\simeq
3150$~m/s and coherence lives considerably longer.

\begin{figure}[t]
\begin{center}
\epsfig{file=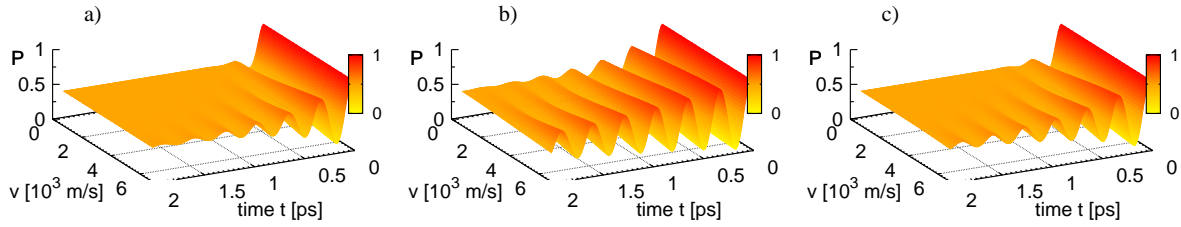,width=16cm}
\end{center}
\caption{\label{fig4}
 Occupation difference of donor and acceptor
versus time and speed of sound assuming propagating environmental
modes. Parameters are chosen to be $\omega_c=\Delta=106$~cm$^{-1}$
and $r_{da}=3.8$~\AA, and in a) $T=152$~K, $\alpha=0.2$, in b) 
$T=76$~K, $\alpha=0.08$, and in c) $T=304$~K, $\alpha=0.08$.}
\end{figure}

In order to elucidate the dependence of this effect on the system-bath coupling, 
Fig.\ \ref{fig4} a) shows the result for $\alpha=0.2$, resulting in a
reorganization energy $\lambda\simeq 2\alpha\omega_c=42.5$~cm$^{-1}$.
This 
compares to the case shown in Fig.\ \ref{fig2}, where 
we have set $\alpha=0.08$, resulting in a
reorganization energy $\lambda\simeq 2\alpha\omega_c=17$~cm$^{-1}$.
The
stronger coupling results in faster decoherence and the crossover to
fully correlated environmental fluctuations causing long-time
coherence is pushed to larger values of the sound velocity.

\begin{figure}[t]
\begin{center}
\epsfig{file=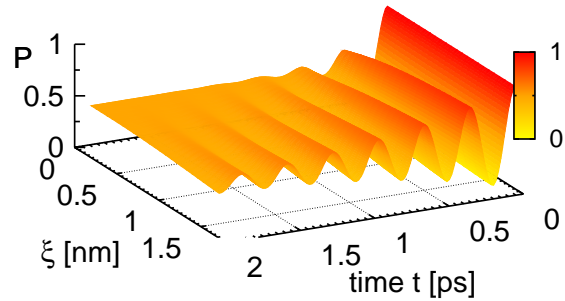,width=8cm}
\end{center}
\caption{\label{fig3} Occupation difference of donor and acceptor
versus time and localization length $\xi$ of localized environmental modes.
Parameters are chosen as $\omega_c=\Delta=106$~cm$^{-1}$,
$T=152$~K, $\alpha=0.08$ and $r_{da}=3.8$~\AA.}
\end{figure}

To address the temperature dependence, we show in 
Fig.\ \ref{fig4} b)
and c) the results for $P(t)$ for the same parameters as 
in Fig.\ \ref{fig2}, except that 
temperature is set to $T=76$~K in b) and $T=304$~K in c). At lower
temperatures coherence expectedly survives longer whereas 
at higher temperatures coherence lives shorter. 
Nevertheless, for all temperatures the
profound effect due to the finite sound velocity is present. 
This turnover between uncorrelated (strong decoherence) and strongly
correlated (weak decoherence) environmental fluctuations thus is only
weakly dependent on temperature.
Next, we discuss the dependence of the crossover on the localization 
length $\xi$. Fig.~\ref{fig3} shows $P(t)$ for localized modes with localization 
lengths between $0$ - $2$~nm. Again as expected, for small localization
lengths, the fluctuations at each site are uncorrelated and the
occupation difference decays in less than a picosecond. Assuming a
distance between donor and acceptor of $r_{12}=3.8$~\AA, the
localization lengths plotted in Fig.~\ref{fig3}, reach up to about 
four times the donor-acceptor distance.
Then sizable correlations are expected and coherent
oscillations for more than 2 picoseconds occur. 
In conclusion, assuming localized or propagating modes results
qualitatively in the same behavior. The used parameters are all taken
for the case of the FMO complex and thus our results strongly suggest that
spatial correlations of the environmental fluctuations due 
to finite propagation time of the modes strongly
influence the decay of coherence in exciton transfer processes. In
order to judge on the quantitative effect, a comprehensive experimental
investigation of the environmental modes is needed, in particular, 
 whether the modes are propagating or localized, and accordingly, 
whether the sound velocity and/or the localization
length is relevant.
So far, we have discussed the influence of spatially correlated
environmental fluctuations on the coherence of a {\em single\/} 
donor-acceptor pair.
These results can in principle be extended to a chain of 
more chromophoric sites without changing the 
physical picture qualitatively.

\section{Transfer in two chromophore pairs}

\subsection{Model}

The FMO complex consists of three identical subunits, each of which
consists of seven chromophoric sites and acts as a conductor for the
excitons. Most likely, this structure has been optimized with respect 
to efficiency and seems to contain also some redundancy, which 
might be a measure of reliability in
nature. In any case, the complex structure gives raise to the question 
whether a crosstalk between the subunits exists, and this even in 
a quantum coherent manner. In turn, the question whether this serves 
any purpose for functionality of increased efficiency is reasonable.

In order to approach this question on a qualitative level, simple 
low-dimensional effective models are necessary. We model a single 
subunit by one 
donor-acceptor pair and discuss in the following two such
donor-acceptor pairs located at a distance $r$. We assume that each 
pair initially contains a single exciton at the respective donor site. 
We explicitly suppress exciton transfer from one pair to the other and 
start from the Hamiltonian 
\begin{eqnarray}
\label{Hpda} 
H_{\rm pda} &=& \frac{1}{2}\Delta \sum_{j=1}^2 \left\{
|d_j\rangle\langle a_j| \,+|a_j\rangle\langle d_j| \right\} \,+
\sum_{j=1}^2\sum_{i=a_j/d_j}|i\rangle\langle i| \sum_{\bf k}\lambda_{\bf k}({\bf
r}_i) q_{\bf k} \nonumber \\ 
& & +\frac{1}{2}\sum_{\bf k} \left(p_{\bf k}^2+\omega_{\bf k}^2
q_{\bf k}^2 \right) \, .
\end{eqnarray}
We assume in the following that 
each chromophore couples separately to the environmental fluctuations 
but in contrast to the previous section, here there are two distinct
distances involved. As before, each donor is separated by a distance
$r_{da}=|{\bf r}_{a_1}-{\bf r}_{d_1}|=|{\bf r}_{a_2}-{\bf r}_{d_2}|$ 
from its acceptor. We assume this distance to be identical
for both donor-acceptor pairs. The separation between both pairs is
$r=|{\bf r}_{a_1}-{\bf r}_{a_2}|=|{\bf r}_{d_1}-{\bf r}_{d_2}|$.

\subsection{Results}

\begin{figure}[t]
\begin{center}
\epsfig{file= 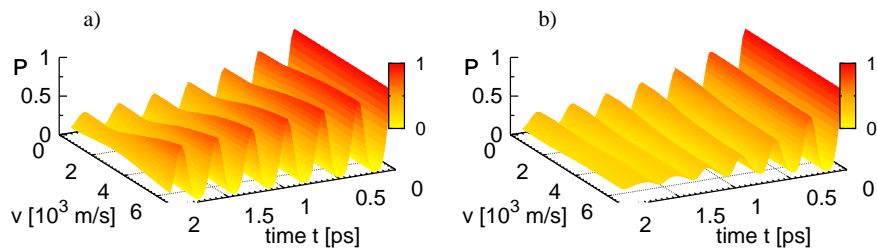,width=12cm}
\end{center}
\caption{\label{fig8}Probability $P_{d_1,d_2}(t)$ of both exciton
being at the 
respective donor sites versus time and sound velocity assuming
propagating environmental modes. Parameters are chosen to be
$\omega_c=\Delta=106$~cm$^{-1}$, $T=15.2$~K, and $\alpha=0.04$. 
The distances are for a) $r_{da}=3.8$~\AA~and $r=38$~\AA~, and for b) $r_{da}=38$~\AA~and $r=3.8$~\AA.}
\end{figure}

Since we have seen that propagating and localized (with a finite
localization length) 
modes cause similar results for the decay of coherence, we restrict
the following investigation to propagating modes. As before we set the
tunneling element $\Delta=106$~cm$^{-1}$ and the  
fluctuation cut-off frequency $\omega_c=106$~cm$^{-1}$.
Fig.~\ref{fig8} shows the probability $P_{d_1,d_2}(t)$ that both
excitons are located 
at the donor sites versus time for various values of the sound
velocity for a weak coupling $\alpha=0.04$. Note that we start from
$P_{d_1,d_2}(t=0)=1$. The reorganization energy is then 
$\lambda\simeq 2\alpha\omega_c=8.5$~cm$^{-1}$ and a rather low
temperature $T=15.2$~K is chosen. In Fig.~\ref{fig8} a), we fix the
distances $r_{da}=3.8$~\AA~and $r=38$~\AA~which reflects two rather
distant donor-acceptor pairs. Both donor-acceptor pairs should thus 
be independent of each other since no direct coupling is assumed. We 
find a qualitatively similar behavior as for the case of a single pair
as described in Section \ref{single}. Differences arise due to the
weaker coupling and lower temperature. In contrast to the former
case, 
we show the results for two close-by donor-acceptor pairs, 
$r=3.8$~\AA~but with large donor-acceptor distance $r_{da}=38$~\AA~  
 in Fig.~\ref{fig8} b). Here, we find that the effect of decoherence 
increases with increasing sound velocity, in clear contrast 
to the previous discussion in Section \ref{single}. Hence, increasing 
spatial correlations between the two donor-acceptor pairs 
destroy quantum coherence.

\begin{figure}[t]
\begin{center}
\epsfig{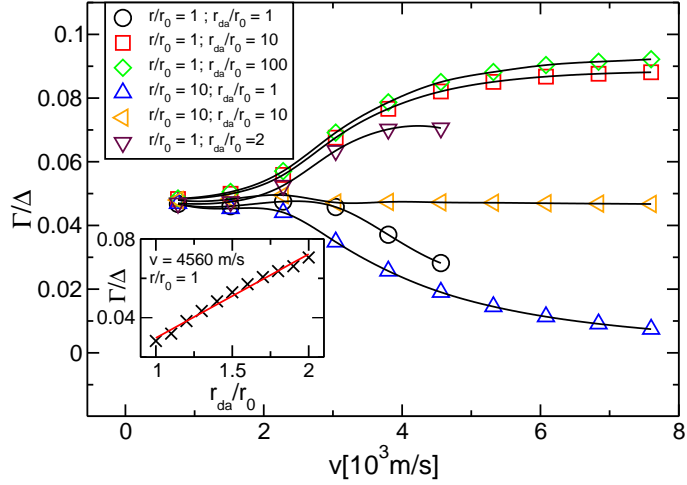}
\end{center}
\caption{\label{fig6} Decoherence rate $\Gamma$ associated to the occupation probability $P_{d_1,d_2}(t)$  versus sound velocity $v$. Parameters are chosen as $\Delta=106$~cm$^{-1}$, $T=15.2$~K, $\alpha=0.04$ and $\omega_c=106$~cm$^{-1}$. Inset: $\Gamma$ versus 
the ratio $r_{da}/r_0$ for $v=4560$~m/s in the range $r_0\le r_{da}\le 2r_0$.}
\end{figure}

For a quantitative investigation of this observation, we determine 
the associated decoherence rate $\Gamma$ by fitting an exponentially
damped 
cosine to the data for $P_{d_1,d_2}(t)$ shown in Fig.~\ref{fig8} 
and plot it in Fig.~\ref{fig6} versus the sound velocity for several 
distance ratios. 
As fundamental distance scale, we use $r_0=3.8$~\AA, the distance of 
chromophore $1$ and $2$ in the FMO complex of {\it Chlorobium
tepidum}~\cite{Bio4}. At the sound velocity $v=7600$~m/s, a mode can
travel this distance within the time of $\Delta^{-1}=50$~fs.

When both donor-acceptor pairs are far apart, $r=10\,r_0$, and the 
distance between donor and acceptor is $r_{da}=r_0$ (shown by the blue
up-triangle in Fig.~\ref{fig6}), we recover the result of the previous
Section \ref{single}. The decoherence rate decreases with increasing
sound velocity. When the donor-acceptor distance is also large,
$r_{da}=10\,r_0$, there is no dependence of $\Gamma$ on $v$ 
for the investigated range of sound velocities (yellow left triangles 
in Fig.~\ref{fig6}). The spatial correlations of the fluctuations
simply do not extend from the donor to the acceptor site and thus the
fluctuations are uncorrelated. When the two pairs are close to each
other and the  donor and acceptor sites are also close, the
decoherence rate $\Gamma$ decreases with increasing sound velocity
(black circles in Fig.~\ref{fig6}). A totally different case is
reached when donor and acceptor sites are well separated,
$r_{da}=10\,r_0$ and $r_{da}=100\,r_0$ (red squares and green diamonds
in Fig.~\ref{fig6}), but the two pairs are close, $r=r_0$. With
increasing sound velocity, the decoherence rate increases and
approximately doubles. The effect is slightly larger with larger
distance between donor and acceptor.

With increasing sound velocity, the wave lengths of the propagating
modes increase and thus spatial correlations of these modes reach
further. When the fluctuations at the sites of both donor-acceptor
pairs are fully correlated, which is approximately the case for sound
velocities of $v\approx 7600$~m/s for $r=r_0$, then the fluctuations
of the second donor (acceptor) equally influence the first one. Thus,
the influence of the 
fluctuations effectively doubles which is reflected in doubling the 
decoherence rate. When donor and acceptors are closer, they as well
become more correlated with increasing sound velocity, which, as
discussed in the Section \ref{single}, results in decreasing
decoherence rates. Thus, two effects are competing here. This also
explains why for $r_{da}=10\,r_0$ (red squares in Fig.~\ref{fig6}) the
increase of the rate $\Gamma$ is weaker than for $r_{da}=100\,r_0$
(green diamonds in Fig.~\ref{fig6}). When both distances are equal,
$r_{da}=r=r_0$ (black circles), the suppression of decoherence due to
spatial correlations between donor and acceptor sites is the dominant
effect, but it is weakened in comparison to the case of $r_{da}=r_0$
and $r_{da}=10\,r_0$ (blue up triangles). Hence, we find two regimes,
in which either one of the two effects dominates. 
How sharp the crossover between the two regimes is, becomes visible when looking at the data for $r_{da}=2\,r_0$ and $r=r_0$. Here again, the decoherence rate $\Gamma$ is increased by spatial correlations. Similar behavior (not shown) is also found for a smaller fluctuation cut-off frequency $\omega_c=53$~cm$^{-1}$ and otherwise identical parameters. The inset of Fig.~\ref{fig6} shows the change of the decoherence rate $\Gamma$ when changing the donor-acceptor distance $r_{da}$ from $r_0$ to $2\,r_0$ for a fixed sound velocity  
$v=4560$~m/s. We find that in this regime, the change is linear in $r_{da}$. 

\begin{figure}[t]
\begin{center}
\epsfig{file= 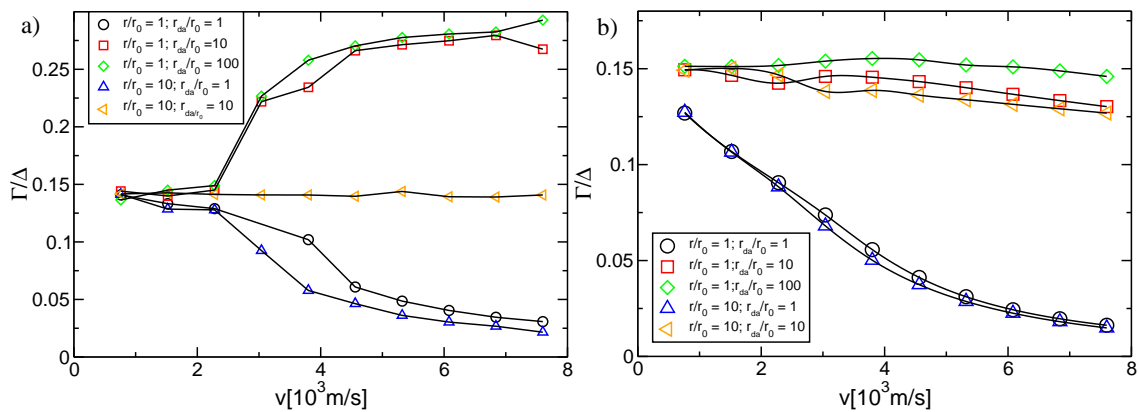,width=15cm}
\end{center}
\caption{\label{fig7} Decoherence rate $\Gamma$ associated to the occupation probability $P_{d_1,d_2}(t)$  versus sound velocity $v$. Parameters are $\Delta=106$~cm$^{-1}$, $\alpha=0.04$, and in a) $T=15.2$~K, $\omega_c=1060$~cm$^{-1}$, and in b) $T=152$~K, $\omega_c=106$~cm$^{-1}$.}
\end{figure}

Qualitatively the same happens for very large fluctuation cut-off frequencies $\omega_c=1060$~cm$^{-1}$ 
(as shown in Fig.~\ref{fig7}a for otherwise identical parameters). Note that in this regime, non-Markovian effects are suppressed. 
The increase (decrease) of $\Gamma$ for correlated donor-acceptor pairs (for correlated fluctuations at donor and acceptor site) 
becomes sharper. 
The influence of temperature is studied in Fig.~\ref{fig7}b), which shows the decoherence rate $\Gamma$ for 
a higher temperature, $T=152$~K, but otherwise for the same parameters as in Fig.~\ref{fig6}. 
For small donor-acceptor distance, $r_{da}=r_0$, we find as before that the decoherence rate decreases with increasing sound velocity. 
However, the picture changes for large donor-acceptor distance, $r_{da}=10r_0$. Then, the effect of decoherence is almost independent 
of the sound velocity, irrespective of the distance $r$ between the donor-acceptor pairs. 

\section{Discussion and conclusions}

To summarize, we have investigated the effect of spatially correlated
environmental fluctuations on the quantum coherent transfer dynamics of
excitons in donor-acceptor systems. 
Technically, spatially correlated environmental fluctuations can be
included by adopting the numerical quasiadiabatic propagator path
integral scheme. The fluctuations at each chromophore 
site generate a separate term in the Feynman-Vernon 
influence phase. The spatial correlations
 generate additional terms which then describe 
 bath modes at different positions. Nevertheless, only the
Feynman-Vernon influence phase is modified and the general QUAPI
approach still remains feasible.

When the spatial correlations of the environmental fluctuations extend
from the donor to the acceptor site, their decohering influence is strongly
reduced since then, the energies at both sites are identically modified.
The energy difference between donor and acceptor is not changed but
only the global reference energy fluctuates which does not influence the
dynamics. The spatial correlations of propagating modes are characterized 
by their wave length which itself is determined by the sound velocity 
assuming linear dispersion. In contrast, the spatial
correlations of localized modes are determined by 
their localization length. Although these two cases are in
principle different, the qualitative effect on
quantum coherence is the same and depends only on the ratio of distance
between donor and acceptor and the correlation length given either by
the localization length or by the wave length of resonant modes.

Two donor-acceptor systems in close proximity in addition show an
increase in the decoherence rate in dependence on their spatial 
distance. When the donor
and acceptor sites are far apart and their fluctuations are only weakly
correlated, correlations of close-by donor-acceptor pairs become relevant
when the distance between the two donor-acceptor systems becomes 
small enough that their
environmental fluctuations are correlated. Then, each donor-acceptor
system ``sees'' the fluctuations at the site of the other one and 
thus the decoherence rate is doubled. For
intermediate spatial distances, both effects are competing with each other.
The effect of an increased decoherence rate, however, is suppressed at 
higher temperatures and is probably less relevant at room temperature.

Our results show that quantum coherence in the excitation transfer crucially
depends on spatial correlations in the environmental fluctuations as
soon as correlations lengths are of the order of the spatial distances of
the chromophores. We have shown that for realistic material parameters (in particular, 
sound velocities), noticeable influence of the finite propagation times of 
environmental modes occurs. In fact, correlated fluctuations can 
reduce the decohering effect of the chromophore environment, 
 when the correlation range extends over typical
transfer distances within a chromophore chain. This effect even survives
(even though diminished) up to room temperature and  
thus might be relevant for exciton transfer in biological systems. These 
findings are in line with the experimental findings~\cite{Bio5,Fleming09a}.

At the same time, correlations between different chromophore chains
increase the decoherence rate and thus a close packing would be 
disadvantageous. This increase is, however, suppressed at room temperature. 
Thus,  increased thermal fluctuations actually indirectly support quantum 
coherence since they reduce spatial correlations which would increase 
decoherence. This effect could be tested experimentally 
by performing the reported electronic 
coherence photon echo experiments at even lower temperature 
at a full FMO complex with all three subunits. 
At low temperature, spatial correlations between the 
FMO subunits should reduce quantum coherence 
more effectively than at higher temperatures.

\ack This work was supported by the Excellence Initiative of the
German Federal and State Governments.

\end{document}